\documentclass{Interspeech}
\usepackage{subcaption}

\interspeechcameraready 

\title{Layer-Wise Decision Fusion for Fake Audio Detection Using XLS-R}

\author[affiliation={}]{Yixuan}{Xiao}
\author[affiliation={}]{Ngoc Thang}{Vu}

\affiliation[nocounter]{Institute for Natural Language Processing}{University of Stuttgart}{Germany}

\email{yixuan.xiao@ims.uni-stuttgart.de, thangvu@ims.uni-stuttgart.de}

\usepackage{comment}

\begin{document}

\maketitle

\begin{abstract}
    
    Recent fake audio detection methods often leverage large speech models to achieve robust speech representations. 
These models are typically very deep, providing multiple layer-wise representations. 
However, current works often rely solely on single layer representation or feature fusion to extract one utterance-level representation for decision making. 
These methods risk underutilizing rich information from multiple layers and might induce feature collapse. 
We propose a novel layer-wise decision fusion method that applies fusion after per-layer decision making and achieves the best cross-dataset performance on In-the-Wild dataset (EER 6.90\%) compared to other strong baselines. 
Our model design also makes the model more transparent, allowing us to conduct detailed analysis to reveal the underlying mechanism of decision making.

\end{abstract}

\section{Introduction}
\label{sec:intro}

Fake audio detection (FAD) refers to the task of determining whether a given speech recording is genuine or synthetically generated (e.g., via Text-to-Speech (TTS) or Voice Conversion (VC)). 
While modern FAD systems can achieve high in-domain performance, a main challenge lies in their generalization abilities \cite{DBLP:conf/interspeech/MullerCDFB22}. 
Specifically, the performance of models trained on one dataset often drop significantly when evaluated on another dataset that may have different synthesis algorithms or recording conditions. 

Emerging speech synthesis methods are being rapidly introduced, and their underlying algorithms show great diversity. As a result, binary classification methods that assume the fake class maintains a stable and representative distribution are inadequate. To address this limitation, recent studies have focused on one-class learning methods \cite{DBLP:journals/spl/ZhangJD21, DBLP:journals/corr/abs-2406-16716}, which learn compact representations of real audio and effectively push fake audio away. The underlying assumption is that real audio exhibits greater consistency than fake ones, forming a compact cluster in the feature space. Such methods have shown promising results in robust fake audio detection. 
Utliizing robust speech representations from large speech models such as Whisper\cite{DBLP:conf/icml/RadfordKXBMS23}, XLS-R\cite{DBLP:conf/interspeech/BabuWTLXGSPSPBC22}, and WavLM\cite{DBLP:journals/jstsp/ChenWCWLCLKYXWZ22} is another direction that boost the performance. Several studies that utilize features from these models--trained on massive and diverse data--have shown impressive cross-dataset performance \cite{pascu24_interspeech, martindonas24_interspeech}.

One key attribute of large speech models is their considerable depth. 
Some studies show that each layer emphasizes different features (e.g., acoustic, phonetic, or linguistic properties), and the inherent characteristics and quality of these features changes from one layer to the next as the data passes through the network \cite{DBLP:conf/asru/PasadCL21,DBLP:conf/icassp/PasadSL23}.
As a result, we argue that methods that use only single-layer feature \cite{pascu24_interspeech,DBLP:conf/icassp/WangY23} or fuse features from multiple layers into one \cite{pan24c_interspeech,martindonas24_interspeech} \emph{before} decision making are suboptimal and might induce \emph{feature collapse}. 
For example, if one dataset emphasizes acoustic properties (e.g., exaggerated pitch) while another highlights para-linguistic attributes (e.g., stress or rhythm), training on the former might lead the model to overvalue layers relevant to acoustic properties, resulting in a homogenized representation with reduced generalizability.

To avoid feature collapse, we propose a layer-wise decision method: a classifier is attached to each layer and the final output is derived by fusing all classifier outputs. 
Such design enhances model transparency and enables the following \emph{layer-wise} analysis: (1) we compare the behavior of feature fusion versus decision fusion models and determine whether feature collapse exists. (2) we examine the impact of including silence during training. (3) we explore the role of discrete tokens with higher attention scores in cross-domain performance to find linguistic cues. A discrete token is an integer label assigned by XLS-R that serves as a pseudo-phoneme.

Our \textbf{contribution}\footnote{\url{https://github.com/XIAOYixuan/tomatoDD/tree/interspeech25-layer-wise}} includes: (1) a novel layer-wise and more interpretable decision fusion architecture that achieves the best cross-dataset performance (EER 6.90\%) on the challenging dataset In-the-Wild \cite{DBLP:conf/interspeech/MullerCDFB22}, (2) analysis that studies 
 factors that affect performance, providing insights for future robust FAD design, (3) a discrete token set that might contribute to further works on explainable fake audio detection.

\noindent\textbf{Analysis Findings:} 
 \emph{\textbf{Analysis 1}} shows that feature collapse exists in the feature fusion method, with each run relying on fewer but different layers, while our decision fusion ones activate more layers and consistently highlights early-middle layers.
 \emph{\textbf{Analysis 2}} shows that including silence in training greatly improves in-domain performance but significantly decreases cross-domain performance. All layer, including the deepest ones, receive impact from including silence. Decision making from shallow layers' classifiers is the most being affected. 
 \emph{\textbf{Analysis 3}} finds moderate correlation between \emph{important discrete tokens similarity} and cross-dataset performance. Qualitative analysis shows that different layers focus on different aspects of speech.

\section{Methods}

\subsection{Feature Fusion Methods}

We implement a \textit{Feature Fusion} (FF) method based on \cite{martindonas24_interspeech} for comparative analysis. 
The method first extracts hidden representations $h_l \in \mathbb{R}^{T \times D}$ ($D=1024$ for XLS-R-300M\footnote{\url{https://github.com/facebookresearch/fairseq/blob/main/examples/wav2vec/xlsr/README.md}}) from a pretrained speech model for layer $l$.
Then it learns a weight vector $\gamma \in \mathbb{R}^{L}$ ($L=25$ for XLS-R 300M, 1 feature extractor layer + 24 encoder layers), which is normalized via softmax to yield $\gamma_{\text{prob}}$. 
These weights are used to aggregate 25 $h_l$ features into $e \in \mathbb{R}^{T \times D}$, which is then projected by a linear mapping $W_p \in \mathbb{R}^{D \times d}$ followed by a ReLU activation to produce $e_{\text{proj}} \in \mathbb{R}^{T \times d}$. 

To get the per-frame attention score for time pooling, an attention mechanism is applied by feeding $e_{\text{proj}}$ through two affine transformations ($M_{attn} \in \mathbb{R}^{d \times 256}$ and $M_{head} \in \mathbb{R}^{256 \times H}$, $H$ is number of attention heads and set to 4) with an intermediate ReLU activation. Each head predicts a raw score for each frame; the four raw scores are then aggregated using a log-sum-exp operation, followed by a softmax normalization. 
After time pooling, the utterance-level representation $f \in \mathbb{R}^{d}$ is used in a one-class softmax classifier \cite{DBLP:journals/spl/ZhangJD21}.

\subsection{Layer-wise Methods}

\begin{figure}[ht]
    \centering
    \includegraphics[width=0.8\linewidth]{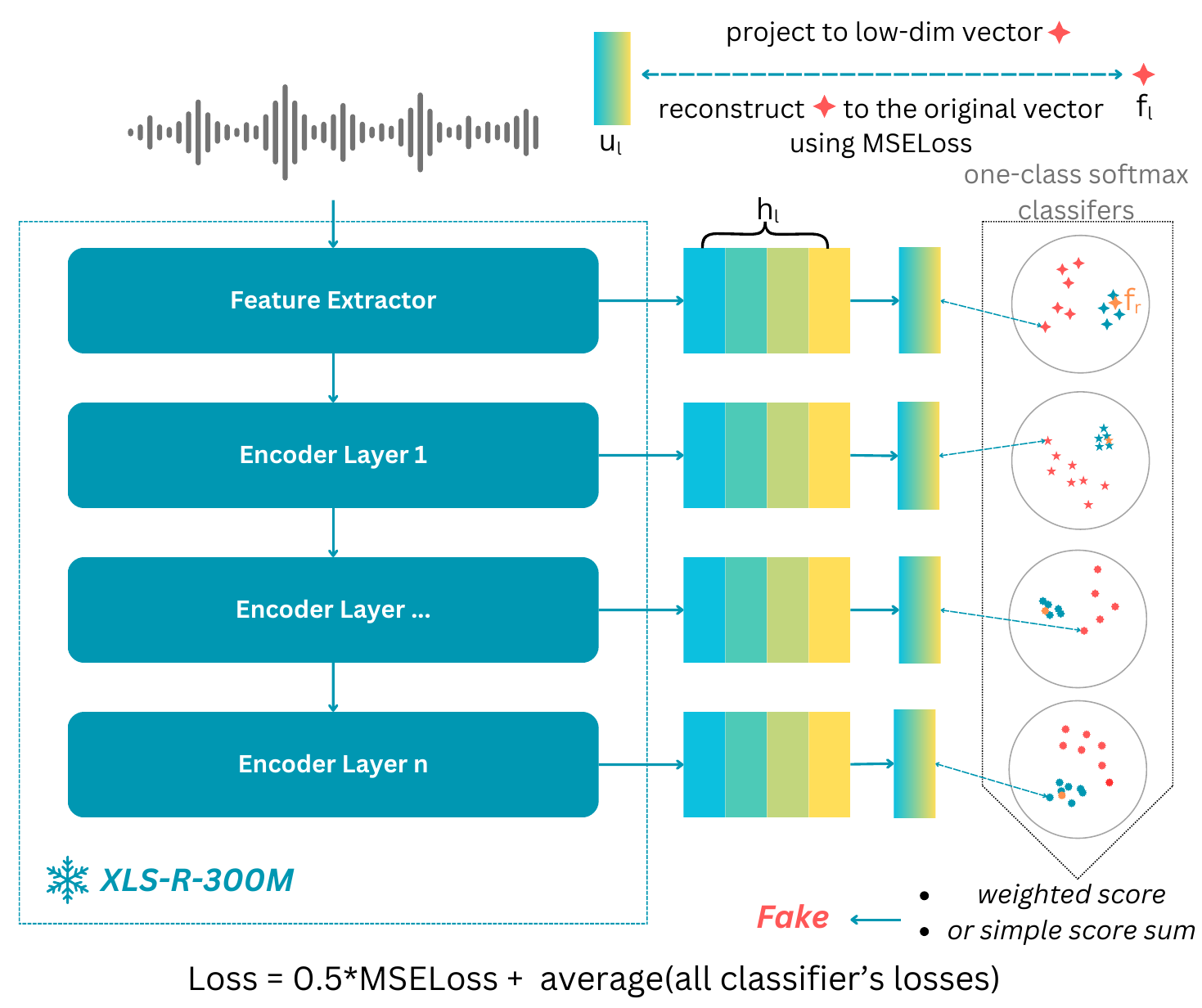}
    \caption{Model architecture.}
    \label{fig:model}
\end{figure}

As shown in Figure~\ref{fig:model}, each layer participates in determining a decision boundary, thereby preserving the unique information captured at different depths. 

For layer $l$, a hidden feature sequence $h_l \in \mathbb{R}^{T \times D}$ is extracted. The method adopts a similar time pooling method to FF. 
We only use one affine transformation $M_{head} \in \mathbb{R}^{D \times H}$ for time pooling, otherwise there will be 25 $M_{attn} \in \mathbb{R}^{D \times 256}$ parameters for a relatively small training set\footnote{Preliminary experiments show that this setting leads to overfitting.
}. 
After time pooling, we achieve a per-layer representation $u_l \in \mathbb{R}^{D}$. 
Similar to FF, to avoid redundancy, $u_l$ can be reduced to a lower-dimensional space $f_l \in \mathbb{R}^{d}$
using a shared bottleneck projection matrix $W_b \in \mathbb{R}^{D \times d}$. 
Inspired by Autoencoder \cite{DBLP:journals/corr/abs-2003-05991}, to prevent the shared projection from discarding important information, a reconstruction loss is used as a regularization term. This model is referred to as \textit{Layer-wise method with Bottleneck Final and Reconstruction loss} (LW\_BNR). If the reconstruction loss is not used, the model is referred to as \textit{Layer-wise method with Bottleneck Final} (LW\_BN). If dimensionality reduction is not applied, the model is simply referred to as \textit{Layer-wise} (LW).

Finally, each layer is attached with a one-class softmax classifier that learns a \textit{real center} $f_r \in \mathbb{R}^{d}$. Each classifier provides a score $s_l$ representing the similarity between the input and $f_r$. Scores from different classifiers are fused in two ways: 1) by summing all scores or 2) by learning a weight for each layer and summing the weighted scores. The latter uses a learnable weight vector $\alpha \in \mathbb{R}^{L}$, normalized via softmax. Models using weighted scores are denoted with a "W" suffix (e.g., LW\_BNW).

\section{Experiment Setting}

\noindent\textbf{Dataset} All models are trained on the ASVspoof19 LA\cite{DBLP:journals/csl/WangYTDNESVKLJA20} train set and evaluated on both the ASVspoof19 LA evaluation set and the In-the-Wild (ITW) dataset. The ASVspoof19 LA contains clean, high-quality recordings with spoofed speech generated using TTS and VC methods. In contrast, ITW collects audio from real-world sources, featuring diverse acoustic conditions and spoofing artifacts. 

\noindent\textbf{Data Processing} We randomly sample four-second segments from the audio recordings following \cite{rawnet}. 
During training, data augmentation is applied to improve generalization. 
Specifically, with a probability of 1/3 the audio is convolved with a room impulse response\footnote{\url{https://www.openslr.org/28/}} to simulate reverberation and different recording environments, and with a probability of 1/5 RawBoost \cite{DBLP:conf/icassp/TakKPTE22} noise is added to simulate channel distortions and transmission artifacts. No augmentation is applied during inference. Leading and trailing silence are removed from the audio. Features are extracted from a frozen XLS-R using Fairseq \footnote{\url{https://github.com/facebookresearch/fairseq/}}.

\noindent\textbf{Training} Models are trained on a 48G RTX A6000 using Adam optimizer with $\beta_1=0.9$, $\beta_2=0.999$, weight\_decay is set to $0.001$. A step scheduler is used to halve the learning rate every 20 epochs. The maximum training epoch is set to 100, and early stopping is applied with a patience of 10 epochs. The batch size is set to 128. Weighted loss (fake:real=1:9) is used to handle class imbalance. 
We train the model five times with different random seeds and report the mean and standard deviation for the results. 

\section{Results and Discussion}

\begin{table}[ht]
    \centering
    \caption{Performance comparison.}
    \begin{tabular}{lcc}
    \toprule
    \textbf{Method} & \textbf{ASV} & \textbf{ITW} \\
    \midrule
    \multicolumn{3}{l}{\textbf{Baseline Methods}} \\
    XLS-R+logres\cite{pascu24_interspeech}: &  &  \\
    \quad 300m (w silence) & 1.00 & 21.30 \\
    \quad 1b (w silence)  & 1.30 & 18.70 \\
    \quad 2b (w silence)   & 0.60 & 7.20  \\
    wav2vec2+binary classifier \cite{DBLP:conf/icassp/WangY23} & 2.98 & 26.65 \\
    NN-ASP (w silence) \cite{martindonas24_interspeech}              & 0.22 & 11.10 \\
    NN-ASP(w/o silence) \cite{martindonas24_interspeech}            & 5.56 & 9.49  \\
    NN-ACP (w silence) \cite{martindonas24_interspeech}              & \textbf{0.19} & 11.09 \\
    NN-ACP(w/o silence) \cite{martindonas24_interspeech}            & 8.09 & 10.27  \\
    \midrule
    \multicolumn{3}{l}{\textbf{Proposed Methods using xlsr-r-300} w/o silence} \\
    FF                           & 7.02~($\pm$0.51)  & 10.97~($\pm$1.30) \\
    LW                           & 5.97~($\pm$0.19)  & 9.11~($\pm$1.14)  \\
    LW\_BN                       & 5.27~($\pm$0.39)  & \textbf{6.90~($\pm$0.30)}  \\
    LW\_BNR                     & 5.08~($\pm$0.13)  & 7.40~($\pm$0.24)  \\
    LW\_BNW             & \emph{\textbf{4.88~($\pm$0.15)}}  & 7.52~($\pm$0.14)  \\
    LW\_BN (w silence)              & 0.33~($\pm$0.02)  & 16.83~($\pm$0.90) \\
    \bottomrule
    \end{tabular}
    \label{tab:results}
\end{table}

Table \ref{tab:results} shows the Equal Error Rate (EER) comparison on the in-domain evaluation set ASVspoof19 LA (ASV) and the out-of-domain evaluation set ITW. Our results are presented as mean~($\pm$ standard deviation) over five runs. We compare our proposed methods with several baseline methods that adopt rather single-layer representation or feature fusion to extract features. All models use frozen large speech models. 

\cite{pascu24_interspeech} and \cite{DBLP:conf/icassp/WangY23} use a single-layer representation and in general show poor cross-dataset performance, showing the limitation of using only one layer for detection. 
However, as the model capacity increases (e.g., from 300M to 2B parameter models used in \cite{pascu24_interspeech}), performance improves noticeably. 
Our proposed methods, along with two baselines \textit{NN-ASP} and \textit{NN-ACP}, which use a 300M parameter model, offers a more memory-efficient alternative: all of them achieve better cross-datasetperformance than the other baselines, except for the largest 2B model.

However, there is still a noticeable performance gap on ASV between our methods and many other baselines. We attribute this to the training setting regarding silence. When the models are trained with silence (not trimming the leading and trailing silence), they achieve very low EERs on ASV; however, the cross-dataset performance on the ITW dataset is poor. We also trained a model (LW\_BN) with silence, and achieved similar results. When trained without silence, our model also achieved the best performance on ASV.
This observation, which aligns with previous studies showing that the silent part exists shortcuts in the official challenge dataset \cite{DBLP:journals/corr/abs-2106-12914}, therefore we believe training with silence can reduces cross-domain generalizability and does not reflect the real-world scenario. In the following, we focus on training without silence.

In comparing our models with \textit{NN-ASP} and \textit{NN-ACP}, we note that \textit{FF} is very similar to \textit{NN-ASP}. The performance gap between these methods likely stems from that our implementation uses a fixed-length input, while \textit{NN-ASP} uses a variable-length input. Since more information is available in the full audio, we expected that their performance is better. Nonetheless, almost all layer-wise decision fusion (\textit{LW}) variants consistently outperform both \textit{NN-ASP} and \textit{NN-ACP}, highlighting the advantage of making decisions using different features and then fusing the decision later.

Among the \textit{LW} variants, \textit{LW\_BN}, \textit{LW\_BNR} and \textit{LW\_BNW} outperform the base \textit{LW} on both datasets. This suggests that projecting the original 1024-dimensional features into a lower-dimensional space may reduce redundancy or simplify the task for the one-class classifier to identify a robust center.
Incorporating a reconstruction loss (LW\_BNR) and a learnable weight for score fusion (LW\_BNW) both boost in-domain (ASV) performance but worsen cross-dataset (ITW) results, reflecting the trade-off between specialized in-domain optimization and broader cross-domain generalization.

\section{Analysis}

In this section, we aim to understand the underlying reasons why LW variants outperform FF. Additionally, we conduct a layer-wise analysis to examine how silence affects performance. Does it only impact shallow layers? Are deeper layers, which are less relevant to simple signal characteristics, more resistant? Furthermore, since we can extract discrete tokens acting like frame labels from a self-supervised model, with the help of our attentive time pooling, can we identify any ``important'' discrete tokens that contribute to cross-dataset performance, which could be used to build a robust model in the future? Also, what linguistic cues can we infer from these potential important tokens? To answer these questions, we conduct three analyses.

\subsection{What causes the difference between FF and LW variants?}

\begin{figure}[ht]
    \centering
    \begin{subfigure}[b]{\linewidth}
        \centering
        \includegraphics[width=\linewidth]{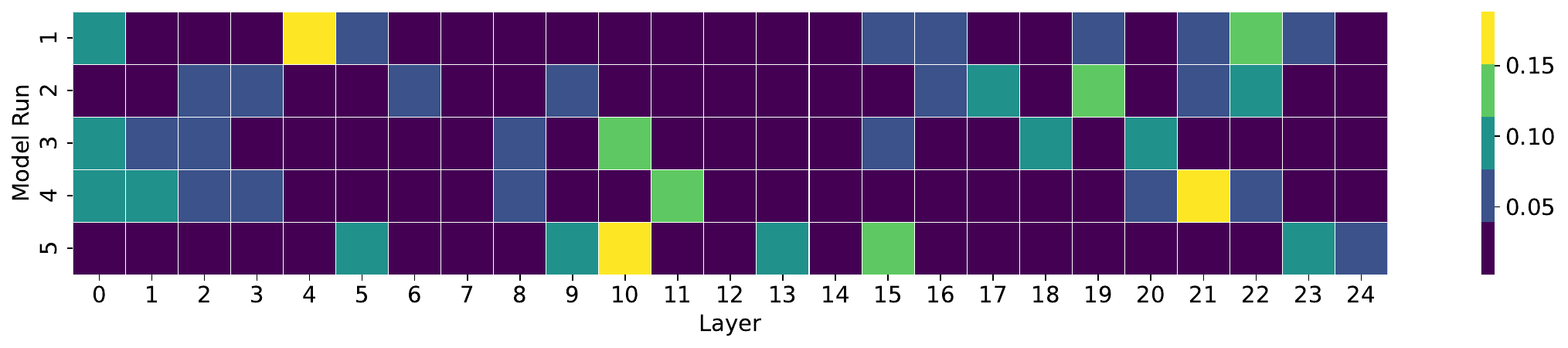}
        \caption{Layer weight heatmap for FF model}
        \label{fig:weight_ff}
    \end{subfigure}
    
    \begin{subfigure}[b]{\linewidth}
        \centering
        \includegraphics[width=\linewidth]{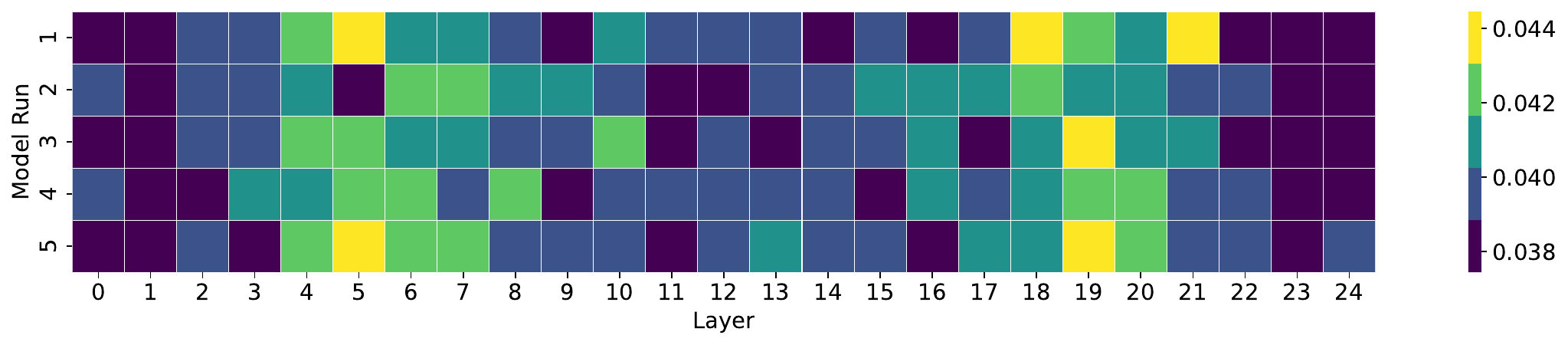}
        \caption{Layer weight heatmap for LW\_BNW model}
        \label{fig:weight_lwbn}
    \end{subfigure}
    \caption{Layer weight heatmap.}
    \label{fig:weight_maps}
\end{figure}

We study the difference by exploring how they use different layers to make decisions.
Figure~\ref{fig:weight_maps} displays the layer weight heatmaps for both FF and LW\_BNW models, where each row corresponds to a different run and the x-axis represents the layer index. 
In the FF model, the layers with higher weights vary across runs, whereas LW\_BNW shows more consistent high weights in layers 4--8 and 18--21, with more layers activated overall. 
This suggests that many layers can provide discriminative features on the in-domain dataset while FF only ``picks'' very few in each run to make a decision, reflecting feature collapse and causing limited generalization. 
In contrast, LW\_BNW uses each layer’s features to determine a decision boundary before fusion, 
preserving more information for decision making. 

\begin{figure}[ht]
    \centering
    \includegraphics[width=\linewidth]{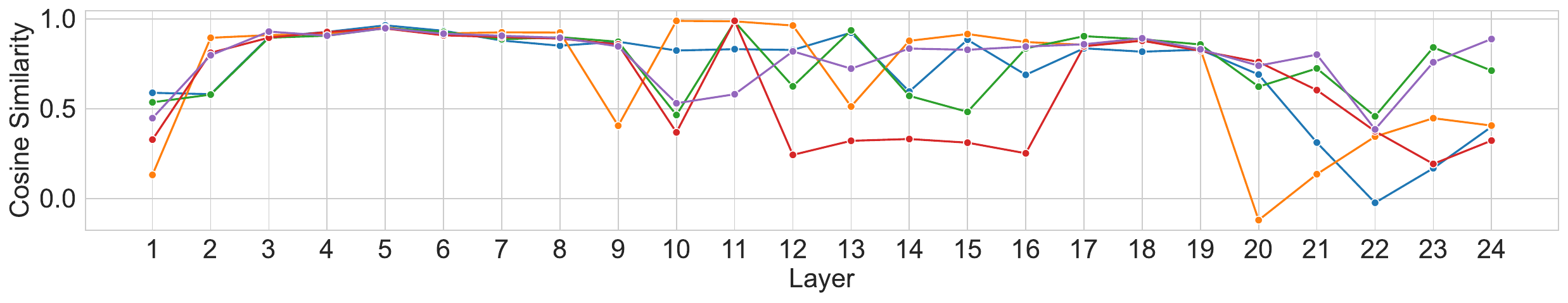}
    \caption{Cosine similarity between consecutive centers.} 
    \label{fig:cosine_centers}
\end{figure}

Although layer weights are not available in LW\_BN, we analyze the consecutive center similarity to study its behavior. In Figure~\ref{fig:cosine_centers}, the value at layer $i$ is the cosine similarity between the real center at layer $i$ and that at layer $i-1$.
Results show that centers in layers 2--8 and 17--19 are highly similar across runs. The layer selection is similar to LW\_BNW. These similar centers can be viewed as the model automatically aggregates and augments important information (by summing their scores up).

\subsection{How does silence affect layer-wise performance?}

\begin{figure*}[t]
    \centering
    \begin{subfigure}[t]{0.45\textwidth}
        \centering
        \includegraphics[width=\linewidth]{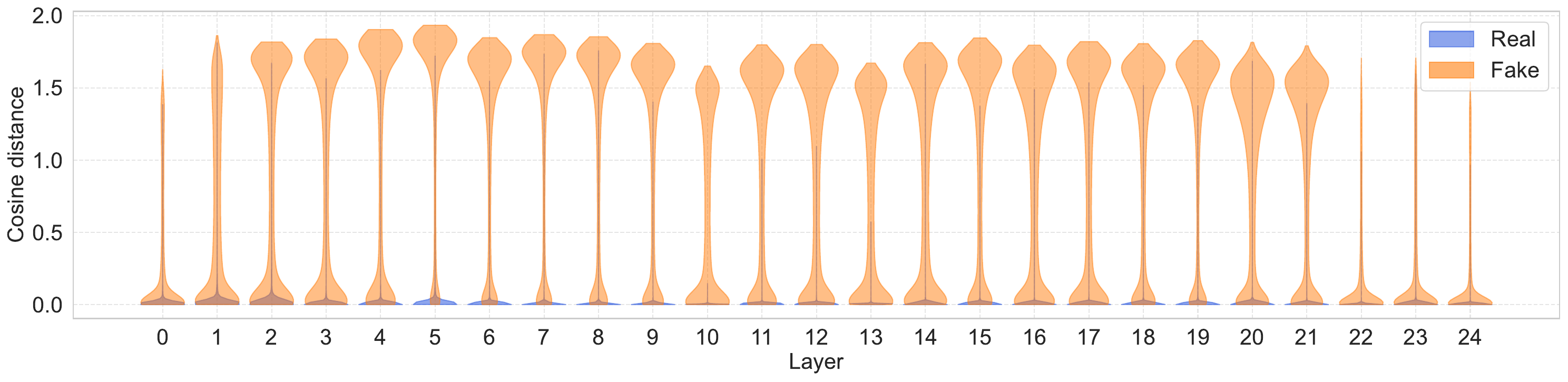}
        \caption{LW\_BN trained without silence: Result on ASV.}
        \label{fig:lwbn_asv}
    \end{subfigure}
    \hfill
    \begin{subfigure}[t]{0.45\textwidth}
        \centering
        \includegraphics[width=\linewidth]{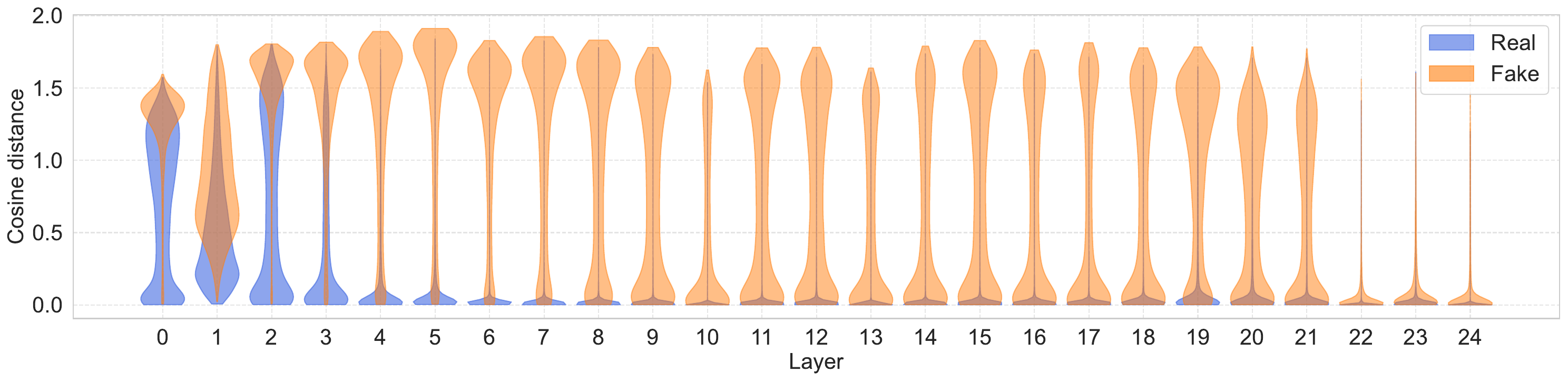}
        \caption{LW\_BN trained without silence: Result on ITW.}
        \label{fig:lwbn_itw}
    \end{subfigure}
    
    \vspace{1em}
    
    \begin{subfigure}[b]{0.45\textwidth}
        \centering
        \includegraphics[width=\linewidth]{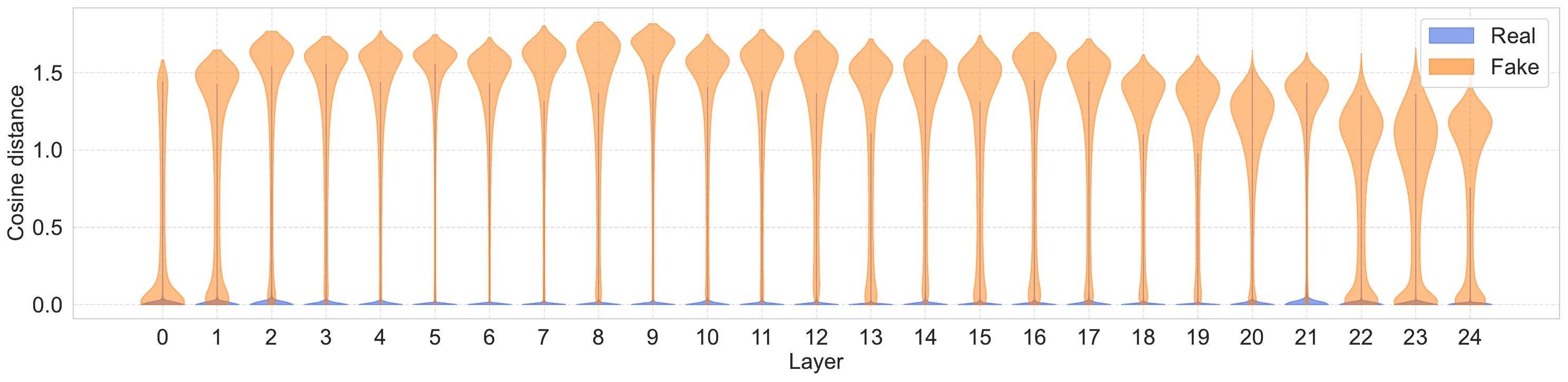}
        \caption{LW\_BN trained with silence: Result on ASV.}
        \label{fig:lwbn_havesil_asv}
    \end{subfigure}
    \hfill
    \begin{subfigure}[b]{0.45\textwidth}
        \centering
        \includegraphics[width=\linewidth]{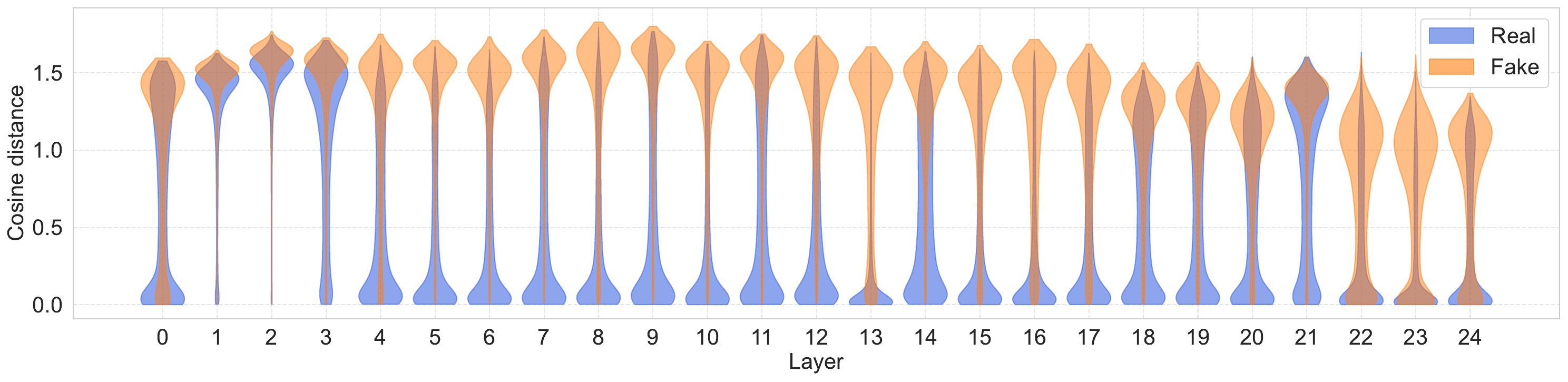}
        \caption{LW\_BN trained with silence: Result ITW.}
        \label{fig:lwbn_havesil_itw}
    \end{subfigure}
    
    \caption{Cosine distance distribution across layers.} 
    \label{fig:distance_dist}
\end{figure*}

We evaluate training with and without silence by comparing the input-to-classifier-center cosine distance distribution for two classes at each layer (see Fig. \ref{fig:distance_dist}). Ideally, fake samples (orange) are far from the classifier center (bottom line) while real samples (blue) remain close.

Figs.~\ref{fig:lwbn_asv} and \ref{fig:lwbn_havesil_asv} show that silent segments clearly carry important information even in the deepest layers for ASVspoof19, since the distance between fake and real samples is larger in \ref{fig:lwbn_havesil_asv}. 
Figs.~\ref{fig:lwbn_itw} and \ref{fig:lwbn_havesil_itw} shows that including silence causes more ITW real samples to be misclassified as fake (especially in shallow layers 1--3), likely because the ITW data's silent parts have different noises and distortions, reducing generalization. When the model is trained without silence, more fake samples are misclassified as real, possibly because: 
1) model trained on ASVspoof19 relies more on real samples' characteristics to make a decision, 
2) ITW's low-quality real samples share less similarity with ASVpsoof19's real samples, hence the model failed to generalize. 

\subsection{Can we find any useful information provided by important discrete tokens to further improve robustness?}

We define \emph{important discrete tokens} of each layer as the tokens 1) with high average attention scores and 2) being ``activated'' repetitively across all trials. 
Discrete tokens can be assigned to each frame by XLS-R, and the frame's attention score is generated by our attention-based time pooling method during inference. Therefore, for a given dataset, token-score pairs can be collected per layer. 
Later, for each layer, by averaging the scores for each token, we can rank the tokens by scores. The top 10\% of tokens of that layer form the \emph{per-trial per-layer important token set}. 
Finally, we find the intersection set across all trials, resulting in the \emph{important token set} for that dataset.

\begin{figure}[ht]
    \centering
    \includegraphics[width=\linewidth]{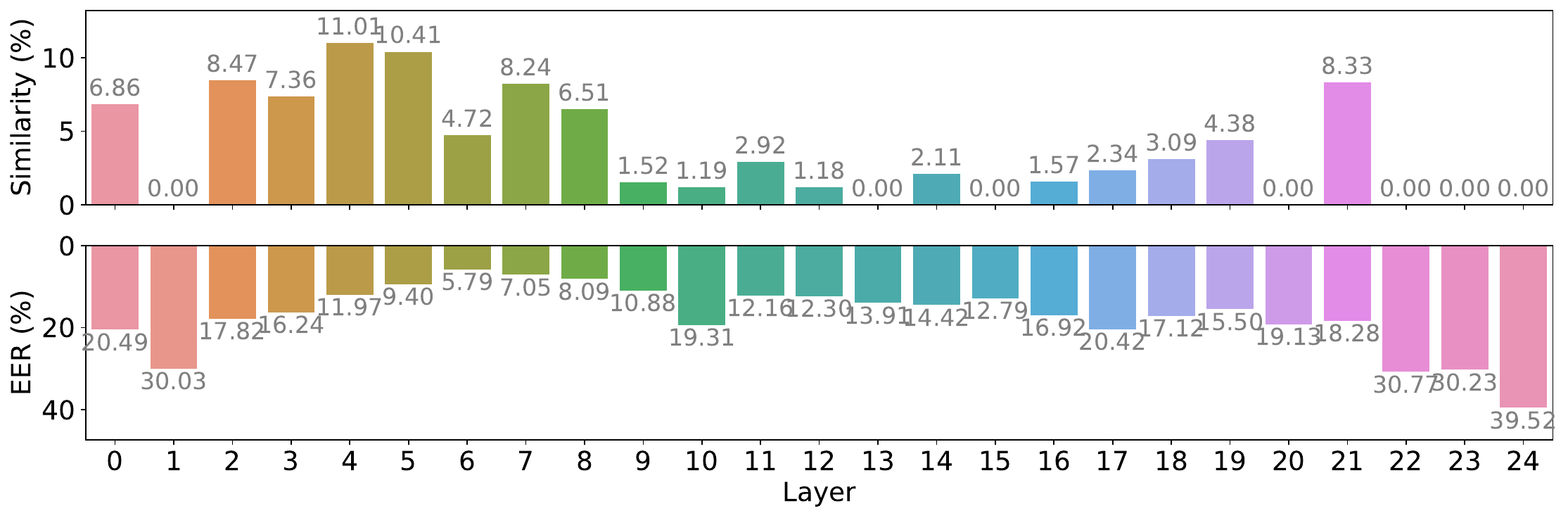}
    \caption{Token set similarity and cross-dataset performance.}
    \label{fig:back_to_back}
\end{figure}

We hypothesize that if two datasets have similar important token sets, the layer-wise classifier has a higher chance to ``hit'' the important tokens on the out-of-domain dataset and achieve better performance. 
To verify this hypothesis, we compare the important token set similarity (measured by Jaccard similarity) and cross-dataset performance (EER) across layers (see Figure~\ref{fig:back_to_back}). However, the correlation is moderate (-0.45, p-value=0.015) and the impact is not consistent. For example, zero similarity can yield very high (Layer 22--24) or fairly good EERs (Layer 13 and 15).

\begin{figure}[ht]
    \centering
    \includegraphics[width=0.9\linewidth]{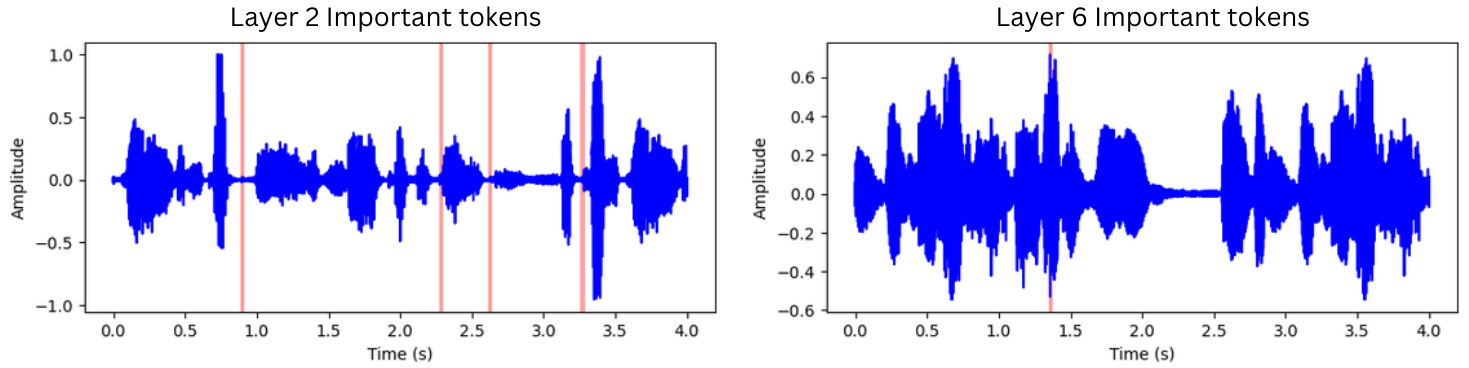}
    \caption{Example audio with highlighted important tokens.}
    \label{fig:audio_example}
\end{figure}

To further explain the inconsistency we observed, we performed a qualitative analysis by focusing on the important frames and their corresponding tokens. 
Any frame with a token label that appears in the important token set is considered important and is highlighted in the waveform. 
We notice that even though some layers have more than 100 important tokens, the number of highlighted frames is much smaller (red area in Figure~\ref{fig:audio_example}). 
Therefore, the majority of the other frames might greatly dilute the impact of important frames after time-pooling. Since whether the other frames might have positive or negative impact is less predictable, the performance is not consistent.

We also extracted the audio segments corresponding to the important frames. After listening, we notice that each layer seems to have its own focus. Figure~\ref{fig:audio_example} shows two examples: audio extracted using the Layer 2 important token set pays more attention to the unvoiced or onset part of the speech, while audio extracted using Layer 6's pays more attention to parts of vowels that sound like /\ae/ or /ei/. However, it is difficult to map the important tokens to specific phonemes. Our observation also aligns with a previous study \cite{DBLP:conf/interspeech/AbdullahSMK23}, which shows for self-supervised models such as XLS-R, phonemes do not map one-to-one with discrete tokens but are rather more related to a distribution over several tokens. A more sophisticated method is needed to further uncover the linguistic cues hidden in the important tokens. As a result, we will also release the important token set under different thresholds to allow future work to further explore the linguistic information.

\section{Conclusion}

We propose a novel decision fusion method that has the best cross-dataset performance compared to baselines. Detailed analyses reveal that, unlike the feature fusion model which suffers from feature collapse and has inconsistent layer reliance, our method includes features from more layers and shows consistent reliance. 
Moreover, including silence has impact on all layers including the deepest ones, although the extent of this impact varies. Finally, cross-dataset important tokens similarity moderately correlates with performance, qualitative analysis shows different layers highlight different aspects of speech.

\bibliographystyle{IEEEtran}
\bibliography{mybib}

\end{document}